\documentclass[11pt]{article}
\usepackage{latexsym}
\usepackage{times}
\usepackage{amssymb}
\usepackage{amsfonts}
\usepackage{amsmath}
\usepackage{amsthm}
\usepackage{srcltx}
\newcommand\F{\mbox{I\kern-2pt F}}

\newcommand\cE{{\cal E}}

\newcommand\cF{{\cal F}}

\newtheorem{theorem}{Theorem}[section]
\newtheorem{proposition}[theorem]{Proposition}
\newtheorem{lemma}[theorem]{Lemma}

\newtheorem{remark}[theorem]{Remark}
\def\E{{\bf E}}
\def\P{{\bf P}}

\def\L{{\bf L}}
\def\Chi{{\bf 1}}
\def\d{\mathrm{d}}
\def\build #1_#2{\mathrel{\mathop{\kern 0pt #1}\limits_{#2}}}
\newcommand{\zs}[1]{{\mathchoice{#1}{#1}{\lower.25ex
\hbox{$\scriptstyle#1$}}
{\lower0.25ex\hbox{$\scriptscriptstyle#1$}}}}

\renewcommand{\theequation}{\thesection.\arabic{equation}}

\def\endproof{\mbox{\ $\qed$}}

\newcommand\bea{\begin{eqnarray}}
\newcommand\eea{\end{eqnarray}}
\newcommand\bean{\begin{eqnarray*}}
\newcommand\eean{\end{eqnarray*}}
\def\bbr{{\mathbb R}}
\def\bbn{{\mathbb N}}

\def\d{\mathrm{d}}

\begin{document}
\title{Ruin probability in the presence of risky
investments.\thanks{This work is partly supported by 
the German Science Foundation (Deutsche 
Forschungsgemeinschaft) through the
Graduiertenkolleg "Angewandte Algorithmische Mathematik" at Munich
University of Technology and by
 RFFI - Grant 04-01-00855. 
 }
}
\author{Serguei Pergamenshchikov
\thanks{Laboratoire de Math\'ematiques Rapha\"el Salem,
UMR 6085 CNRS-Univ. de Rouen
Avenue de l'Universit\'e, BP.12,
76801 Saint Etienne du Rouvray, France,
e-mail:  Serge.Pergamenchtchikov@univ-rouen.fr}
 \and
Omar Zeitouny
\thanks{
Laboratoire de Math\'ematiques Rapha\"el Salem,
UMR 6085 CNRS-Univ. de Rouen
Avenue de l'Universit\'e, BP.12,
76801 Saint Etienne du Rouvray, France,
e-mail:  Omar.Zeitouny@univ-rouen.fr 
 }
}

\date{}
\maketitle

\bigskip

\begin{abstract}
We consider an insurance company in the case when the premium
 rate is a
bounded non-negative 
random function $c_\zs{t}$
and the capital of the insurance company 
is invested in a risky asset
 whose price follows a geometric Brownian
 motion with mean return $a$ and volatility
$\sigma>0$. If $\beta:=2a/\sigma^2-1>0$
we find exact the asymptotic upper and lower bounds for the
ruin probability $\Psi(u)$ as the initial endowment $u$ tends to
infinity, 
i.e. we show that 
$C_*u^{-\beta}\le\Psi(u)\le C^*u^{-\beta}$
for sufficiently large $u$.
 Moreover if $c_\zs{t}=c^*e^{\gamma t}$ with $\gamma\le 0$
 we find the exact asymptotics of the ruin probability, namely
$\Psi(u)\sim u^{-\beta}$. If $\beta\le 0$, we show that
$\Psi(u)=1$ for any $u\ge 0$.
\end{abstract}

\noindent
 {\em MJS:} primary 62P05; 60J25; G22; G23
\medskip


\renewcommand{\theequation}{1.\arabic{equation}} 
\setcounter{equation}{0}
\section{Introduction} 

It is well known that the analysis of activity
of an insurance company in conditions of uncertainty is of 
great importance. Starting from the classical papers of 
Cram\'er  and Lundberg  which 
first considered the ruin problem 
in stochastic environment, 
this subject has attracted much attention. 
Recall that, in the classical Cram\'er--Lundberg model
satisfying the Cram\'er condition and, the positive
safety loading assumption,
the ruin probability as a function of the initial
endowment decreases exponentially (see, for example, Mikosch \cite{Mi}). 
The  problem 
was subsequently
extended to the case when the insurance risk
process is a general L\'evy process
 (see, for example, Kl\"uppelberg et al.
\cite{KlKyMa} for details).

 More recently ruin problems have been studied
in application to an insurance company which
invests its capital in a risky asset
see, e.g., Paulsen \cite{Pa}, 
Kalshnikov and Norberg \cite{KaNo},
Frolova et al. \cite{FrKaPe}
and many others.

 It is clear that, risky investment can be dangerous:
disasters may arrive in the period when the market value of
assets is low and the company will not be able to cover losses by
selling these assets  because of price fluctuations.
Regulators are rather attentive to this issue and impose
stringent constraints on company portfolios. Typically, junk
bonds are prohibited and a prescribed (large) part of the portfolio
should contain non-risky assets (e.g., Treasury bonds) while in
the remaining part only risky assets with good ratings are
allowed. The common notion that investments in an asset 
with stochastic
interest rate  may be too risky for an insurance company can be
justified mathematically.

 We deal with the ruin problem for an insurance company investing
its capital in a risky asset specified by a 
geometric Brownian motion
\begin{equation}\label{1.1}
\d V_t =V_t(a\d t+\sigma\d w_t)\,,
\end{equation}
where $(w_t,\, t\ge 0)$ is a standard Brownion motion and $a> 0$, 
$\sigma>0$.

It turns out that in this case of
{\em small volatility}, i.e.
$0<\sigma^2<2a$, the ruin probability is not exponential but a
power function of the initial capital with the exponent 
$\beta:=2a/\sigma^2-1$. It will be noted that this result holds
without the requirement of positive safety loading.
Also, for large volatility, i.e.
$\sigma^2>2a$, the ruin probability equals $1$ for any 
initial endowment. 
These results 
have been obtained 
 under various conditions in
\cite{Pa,KaNo,FrKaPe}. 

Additionally, a large deviations
limiting theorems 
for describing the ruin probability
was obtained by Djehiche \cite{Dj} and  Nyrhinen \cite{Ny}.
 Gaier et al. \cite{GaGrSc} studied
the optimal investment problem for an insurance company.

In all these papers
the premium rate was assumed to be constant. In practice
this means that the company should 
 obtain a premium with
the same rate continuously. 
We think that this condition is too restrictive 
 and
it significantly bounds the applicability of the above mentioned
results in practical insurance settings.

The goal of this paper is to consider the ruin problem for an
insurance company for which the premium rate is specified by
a bounded non-negative random function $c_\zs{t}$. For the given problem,
under the condition of {\em small volatility}, 
we derive  exact upper and lower bounds for the ruin
probability and in the case of exponential premium rate, i.e.
$c_\zs{t}=e^{\gamma t}$ with $\gamma\le 0$, 
we find the exact asymptotics for the ruin probability.
Particularly, we show that for the zero premium rate, i.e. 
$\gamma=-\infty$, the asymptotic result is
the same as in the case $-\infty<\gamma<0$.

Moreover, in this paper we show that in the  boundary case, i.e.
 $\sigma^2=2a$,
the company goes bankrupt
 with probability $1$ for any bounded function $c_\zs{t}$.

Indeed, an upper bound for the ruin probability
for the random function $c_\zs{t}$ in the small volatility case
is obtained also by Ma and Sun \cite{MaSu}.

 The paper is organised as follows. In the next section we give
the main results. In Section 3 we give the necessary results
about the tails of solutions of some linear random equation
which we apply to study the ruin problem.
In Section 4 we obtain the upper bound for
the ruin probability and in Section 5 we find the corresponding
lower bound.
 In Section 6 we consider the
exponential premium income rate case. In Section 7 we study some ergodic
 properties for an autoregressive process with random coefficient.
in Section 8 we consider
the large volatility case.

\renewcommand{\theequation}{2.\arabic{equation}} 
\setcounter{equation}{0}
\section{Basic results}

Let us consider a process $X=X^u$ of the form
 \begin{equation} \label{2.1}
 X_t=u+  a \int_0^t  X_s  \d s+
  \sigma \int_0^t X_s  \d w_s  + \int_0^t\, c_s \d s  -
\sum_{i=1}^{N_t}\,\xi_i\,,
 \end{equation}
where  $a\ge 0$ and $\sigma\ge 0$ are arbitrary constants,
$w$ is a Brownion motion, $N$ is a Poisson process with
intensity $\alpha>0$ and $(\xi_i\,,i\in\bbn)$ 
are i.i.d. positive
random variables with common a
distribution $F$. Moreover, we assume that  
$w$, $N$, $(\xi_i)$ are independent and
the filtration is defined as 
$\cF_t=\sigma\left\{w_s\,,N_s\,,\sum_{i=1}^{N_s}\,\xi_i\,,
0\le s\le t\right\}$.
Furthermore,
$c_\zs{t}=c(t,X)$ is
a bounded non-negative  $(\cF_t)$ - adapted
function 
(i.e., $0 \leq c_\zs{t}\leq c^*$) such that Eq. \eqref{2.1}
has an unique strong solution 
(see chapter 14 in \cite{Ja}).

 Let $\varsigma_u:=\inf \{t: X^u_t< 0\}$ (the time of ruin),
 $\Psi (u):=P(\varsigma_u<\infty)$ (the ruin probability).
The parameter values $a=0$, $\sigma=0$, $c_\zs{t}=c$, correspond to the
Cram\'er--Lundberg
 model for which  the risk process is usually
written as 
$ X_t=u  + c\,t  -\sum_{i=1}^{N_t}\xi_i$.
 In the considered version (of non-life
insurance) the capital evolves due to a continuously incoming cash
flow with rate $c>0$ and outgoing random payoffs $\xi_i$ at times
forming an independent  Poisson process $N$ with intensity
$\alpha$. For the model with positive safety loading and $F$
having a "non-heavy" tail, the Lundberg inequality provides 
encouraging information: the ruin probability decreases
exponentially as the initial endowment $u$ tends to infinity.
Moreover, for exponentially distributed claims the ruin
probability admits an explicit expression, see \cite{As} or
\cite{Mi}.

We study here the case  $\sigma>0$ with a general random adapted
bounded function $c_\zs{t}$. In this case Eq. \eqref{2.1}
describes the evolution of the  capital of an insurance company,
which is continuously reinvested into an asset with the  price
following a geometric Brownian motion \eqref{1.1}.

Let $\beta:=2a/\sigma^2-1$. 
To write the upper bound for the ruin probability
we define the function :

\begin{equation}\label{2.2}
J(\beta)=\frac{2\alpha}{\sigma^2\beta^2}
\left(
\Chi_\zs{\{0<\beta\le 1\}}+j_1(\beta)\,\Chi_\zs{\{1<\beta\le 2\}}
+j_2(\beta)\,\Chi_\zs{\{\beta>2\}}
\right)\,,
\end{equation}
where $j_1(\beta)=\beta\,(1+\varrho^{-1})$, 
 $j_2(\beta)=
\beta\, 2^{\beta-2}(1+((1+\varrho)^{\frac{1}{\beta-1}}-1)^{1-\beta})$ 
and
$\varrho=\varrho(\beta)=(\beta-1)\sigma^2/2\alpha$.
\begin{theorem}\label{Th.2.1} 
 If $\beta>0$ and   $\E\,\xi_1^\beta<\infty$, then
$\limsup_{u \rightarrow+\infty} u^{\beta}\Psi (u)\le C^*(\beta)$,
where 
$C^*(\beta)= J(\beta)\E\,\xi^\beta_1$.
\end{theorem}
\noindent The proof of this theorem is given in Section 4.

\begin{theorem}\label{Th.2.2}
If $\beta>0$ and
 $\E\xi_1^{\beta+\delta}<\infty$ for  some $\delta>0$,
then there exists a constant $0<C_*<\infty$ such that
 $ \liminf_{u \rightarrow\infty} u^{\beta}\Psi (u)\ge C_*$.
\end{theorem}
\noindent This result is proved in Section 5.
 The following theorem gives the
exact asymptotics for the exponential function $c_\zs{t}$. 

\begin{theorem}\label{Th.2.3} 
 Assume that $c_\zs{t}=c^*\exp\{\gamma t\}$ with 
$-\infty\le\gamma\le 0$. 
If $\beta>0$ and\\ $\E\,\xi_1^{\beta+\delta}<\infty$ 
for  some $\delta>0$,
 then there exists a constant $0<C_\infty<\infty$ such that 
$\lim_{u \rightarrow\infty}\, u^{\beta}\,\Psi (u)= C_\infty$.
Moreover, the constant $C_\infty$ is the same for
any $-\infty\le \gamma<0$.
\end{theorem}

\noindent This result is proved in Section 6. 
Now we consider the large volatility case, 
i.e. $\beta\le 0$.
\begin{theorem}\label{Th.2.4} Assume that
the distribution of 
$\xi_1$ has not a finite support, i.e.  $\P(\xi_1>z)>0$ for any
$z\in\bbr$ .  If $\beta\le 0$ and 
$\E\xi_1^\delta<\infty$ for  some $\delta>0$, then
$\Psi (u)=1$ for any $u\ge 0$.
\end{theorem}
\begin{remark}
This theorem has been proved
by  Paulsen in \cite{Pa}
for a constant premium rate, i.e. for $c_\zs{t}=c^*=\mbox{const}$.
\end{remark}

\noindent The key idea in the proofs of Theorem~\ref{Th.2.1} and
Theorem~\ref{Th.2.2} is based on the fact that 
the function  $\Psi(u)$ may be estimated 
by the tails of solutions of some linear random equations.
 In the next section we study the asymptotic behaviour of those tails.

\renewcommand{\theequation}{3.\arabic{equation}} 
\setcounter{equation}{0}
\section{Tails of solutions of random equations}

This Section contains some results from the general
renewal theory developed by Goldie \cite{Go} for some
random equations. We consider the following two random equations 
\begin{equation}\label{3.1}
R \stackrel{(d)}{=}Q+M\,R\,, \ \ \ \ 
R \ \ \ \ \mbox{is independent of}\ \ \ \ (M,Q)
\end{equation}
($\stackrel{(d)}{=}$ denoting equality of probability laws) and
\begin{equation}\label{3.2}
R^* \stackrel{(d)}{=}Q+M\,(R^*)_+\,, \ \ \ \ 
R^* \ \ \ \ \mbox{independent of}\ \ \ \ (M,Q)\,,
\end{equation}
where $(a)_\zs{+}=\max(a,0)$.

We start with some preliminary conditions for the random variable $M$
which are studied by Goldie (see Lemma 2.2 in \cite{Go}).

\begin{lemma}\label{Le.3.1}
Let $M\ge 0$ be a random variable such that, for  some $\beta>0$
\begin{align}\label{3.3}
\E\,M^\beta\,=1\,, \ \ \ \E\,M^\beta\,(\log\,M)_+\,&<\infty
\end{align}
and the conditional law of $\log\,M$, given $M\ne 0$, 
be non-arithmetic.
Then $-\infty\le \log \E M <0$
and $0<\mu:=\E M^\beta \log M<\infty$.
\end{lemma}

\noindent The following result from \cite{Go}
specifies the tail behaviour of $R$.

\begin{lemma}\label{Le.3.2}(Theorem 4.1 in \cite{Go})
  Let $M$ be a random variable satisfying the conditions of  
Lemma~\ref{Le.3.1} for some $\beta>0$ and  $Q$ be a positive random
variable for which $\E\,Q^\beta<\infty$. Then there is a unique
law for $R$ satisfying \eqref{3.1} such that
\begin{equation}\label{3.7}
\lim_{u\to+\infty}\,u^\beta\,\P(R>u)=c_\infty\,,
\end{equation}
where $c_\infty= \E\,\left((Q+MR)_+^\beta-\,(MR)_+^\beta\right)/\beta\,\mu$
and $\mu=\E\,M^\beta \log\,M$.
\end{lemma}

\noindent Now we study the tail of $R^*$.

\begin{lemma}\label{Le.3.3}
  Let $M\ge 0$ be a random variable satisfying the conditions of  
Lemma~\ref{Le.3.1} for  some $\beta>0$. Assume also that
the distribution of $M$ is absolutely continuous with respect to
Lebesgue measure and there exists $\delta>0$ such that
\begin{equation}\label{3.8}
\E\,M^{\beta+\delta}<\infty 
\end{equation}
and for any $x\in\bbr$
\begin{equation}\label{3.9}
\E\,M^{\beta+\delta+ix}\ne 1\,,
\end{equation}
where $i=\sqrt{-1}$. 
Then under the condition
\begin{equation}\label{3.10}
\E\,|Q|^{\beta+\delta}<\infty
\end{equation}
 for some $\delta>0$
there is a unique
law for $R^*$ satisfying \eqref{3.2} such that there exists
$\lim_\zs{u\to\infty}u^\beta\P(R^*>u)=c^*_\infty$
 and $0<c^*_\infty<\infty$.
\end{lemma}
\noindent This lemma  follows directly
from Theorem 6.3 in \cite{Go}
and Theorem 2 in \cite{Ny}.

\renewcommand{\theequation}{4.\arabic{equation}} 
\setcounter{equation}{0}

\section{Upper bound for the ruin probability}

  Let $\tau_n$ be the instant of
$n$-th  jump of $N$ and let $\theta_n:=\tau_n-\tau_{n-1}$ with
 $\tau_0:=0$. We define the  discrete-time process
$S=S^u$ with $S_n:=X_{\tau_n}$. Since  ruin may occur only when
$X$ jumps downwards,
 $\Psi (u)=P(T_u<\infty)$, where
\begin{equation}\label{4.1} 
T_u:=\inf\{n\ge 1:\ S_n< 0\}.
 \end{equation}
Therefore to obtain asymptotic properties of $T_u$ as 
$u\to\infty$ we need to study the process $(S_n)$. First of all,
we need to find a recurrence equation for this sequence. We start with resolving
of Eq. \eqref{2.1}. For this we introduce the 
 process $(\phi^{s,x}_t)_\zs{t\ge s}$
which satisfies the following stochastic differential equation
$$
\d \phi^{s,x}_t\,=\,a\,\phi^{s,x}_t\,\d t\,+
\,\sigma\,\phi^{s,x}_t\,\d w_t\,+\,c_\zs{t}\,\d t\,, \ \ \
\phi^{s,x}_s=x\,.
$$
The Ito formula implies that 
$\phi^{s,x}_t\,=\,e^{h_t-h_s}\,x\,+\,\int^t_s\,e^{h_t-h_u}\,c_u\,\d u$,
where\\ $h_t=\kappa\,t+\sigma w_t$, 
$\kappa=a-\sigma^2/2$ and $t\ge s$.
 Moreover we can represent Eq. \eqref{2.1} for 
$\tau_{n-1}\le t<\tau_n$ in the following way 
\begin{align*}
X_t&=\,S_{n-1}\,+\, a \int_\zs{\tau_{n-1}}^t  X_s  \d s+
  \sigma \int_\zs{\tau_{n-1}}^t X_s  \d w_s  + 
\int_\zs{\tau_{n-1}}^t\, c_s \d s\,=\,\phi^{\tau_{n-1},S_{n-1}}_t\\
&=\,
\,e^{h_t-h_\zs{\tau_{n-1}}}\,S_{n-1}\,+\,
\int^t_\zs{\tau_{n-1}}\,e^{h_t-h_u}\,c_u\,\d u\,.
\end{align*}
Therefore $S_n=X_\zs{\tau_{n}}=\phi^{\tau_{n-1},S_{n-1}}_\zs{\tau_n}-\xi_n$.
From this we obtain the following random recurrence equation for $(S_n)$ 
\begin{equation}\label{4.2}
 S_n\,=\,\lambda_n\,S_{n-1}\,+\,\zeta_n\,, \ \ \ S_0=u
 \end{equation}
with  $\lambda_n=\exp\{\sigma\,w_{\theta_n}^n+\kappa\theta_n\}$ and 
 $\zeta_n=\eta_n-\xi_n$. Here
$w_{t}^n=w_\zs{t+\tau_\zs{n-1}}- w_\zs{\tau_\zs{n-1}}$ and
$\eta_n=\int_0^{\theta_n}c^n_u
e^{h_\zs{\tau_n}-h_\zs{u+\tau_{n-1}}} \d u$ with
$c^n_u:=c_\zs{u+\tau_{n-1}}$. 
By resolving \eqref{4.2} we find the following representation
for $(S_n)$
\begin{equation}\label{4.3}
 S_n\,=\, \cE_n\,u+\cE_n\sum_{k=1}^n\cE_k^{-1}\,\zeta_k\,, \quad 
\cE_n \,=\, \prod_{k=1}^n\lambda_k\,.
\end{equation}
\noindent Moreover, taking into account here that $\zeta_k\ge -\xi_k$
we obtain that $S_n\ge \cE_n\,(u-Y_n)$,
where
\begin{equation}\label{4.4}
Y_n=Q_1+\sum_{k=2}^{n}\,Q_k\,\prod_{j=1}^{k-1}\,M_j\,, \ \ \
M_j=\lambda_j^{-1}\,, \ \ \ 
Q_k=\xi_k/\lambda_k\,.
\end{equation}
 Notice that 
 $(M_n)$ are i.i.d. random 
variables such that for $q \in ]0,\beta]$
\begin{equation}\label{4.5}
\E\,M_1^q\,=\, \E\,\lambda_1^{-q}=\frac
{2\alpha}{2\alpha+(\beta-q)\,q\,\sigma^2 }\le 1\,.
\end{equation}
 Therefore, there exists $0<\delta<\min(1,\beta)$ for which $\rho=\E\,M^\delta_1<1$
and
$$
\E\,\left(\sum_{k\ge 2}\,Q_k\,\prod_{j=1}^{k-1}\,M_j\right)^\delta\,
\le\,\sum_{k\ge 2}\,\E\,\left(Q_k\,\prod_{j=1}^{k-1}\,M_j\right)^\delta
=
\E\,Q^\delta_1\,
\sum_{k\ge 2}\,\rho^{k-1}\,<\,\infty\,,
$$
i.e. the series 
$\sum_{k\ge 2}\,Q_k\,\prod_{j=1}^{k-1}\,M_j$
is finite a.s. It means that the sequence $(Y_n)$ have a finite limit
\begin{equation}\label{4.6}
\lim_{n\to\infty}\,Y_n\,=\,
Q_1+\sum_{k=2}^{+\infty}\,Q_k\,\prod_{j=1}^{k-1}\,M_j =Y_\infty=R\,<\,\infty
\quad \mbox{a.s.}
\end{equation}
Taking into account that the sequence $(Y_n)$ in (\ref{4.4})
is increasing we can estimate 
$S_n$ as 
\begin{equation}\label{4.7}
S_n\,\ge\,\cE_n\,(u-R)
\end{equation}
and by (\ref{4.1}) we get that $\P(T_u<\infty)\le \P(R>u)$.
Therefore, to obtain the upper bound for the ruin probability
we investigate  the tail behaviour of $R$ as $u\to\infty$. 
To this end, first notice that we may represent $R$ in the following form 
\begin{equation}\label{4.8}
R=Q_1+M_1\,R_1\,,
\end{equation}
where the random variable
$R_1=Q_2+\sum_{k=3}^{+\infty}\,\prod_{j=2}^{k-1}\,M_j\,Q_k$
 has the same distribution as $R$ and is
independent of $(Q_1,M_1)$. Thus the random variable $R$
satisfies Eq. \eqref{3.1}.
 We show that
\begin{equation}\label{4.9}
\lim_{u \rightarrow\infty}\, u^{\beta}\,\P(R>u)= C_1\,,
\end{equation}
where $C_1=2\alpha
\E\,((\xi_1+R)^\beta-R ^\beta)/\beta^2\,\sigma^2$.

To show (\ref{4.9}) we need to check the conditions 
of Lemma~\ref{Le.3.2} for the random
variables $(M_j)$ and $(Q_j)$ defined in (\ref{4.4}). The first property in
(\ref{3.3}) follows directly from (\ref{4.5}) for $q=\beta$.
 Now we show the second. 
By definition of $M_1$ we have 
\begin{align*}
\E\,M^\beta_1\,(\log\,M_1)_+&=
\E\,e^{-\beta\,\sigma\,w_{\theta_1}-\beta\,\kappa\,\theta_1}\,
(-\sigma\,w_{\theta_1}-\kappa\,\theta_1)\,
\Chi_\zs{\{-\sigma\,w_{\theta_1}-\kappa\,\theta_1\ge 0\}}\\
&\le\,\sigma\,
\E\,|w_{\theta_1}|\,
e^{-\beta\,\sigma\,w_{\theta_1}-\beta\,\kappa\,\theta_1}\,
+\,\kappa\,
\E\,\theta_1\,
e^{-\beta\,\sigma\,w_{\theta_1}-\beta\,\kappa\,\theta_1}\,.
\end{align*}
Taking into account that  $(w_t)$ is independent of 
$(\theta_j)$,  the 
last term in this inequality equals 
\begin{align*}
\sigma\,\frac{1}{\sqrt{2\pi}}
\E\,\sqrt{\theta_1}\,
\int^{+\infty}_{-\infty}\,|z|\,
e^{-(z+\beta\,\sigma\,\sqrt{\theta_1})^2/2}\,\d z
+\,\kappa\,\E\,\theta_1\,,
\end{align*}
i.e. $\E\,M^\beta_1\,(\log\,M_1)_+\le\,
(\beta\sigma^2+\kappa)\,\E\,\theta_1
+\sigma\,\sqrt{2/\pi}\,\E\,\sqrt{\theta_1}<\infty$.
In similar way we calculate 
$\mu=\E M^\beta_1 \log\,M_1=\beta\sigma^2/2\alpha$.
Moreover,
$\E\,Q^\beta_1=\E\,\xi^\beta_1<\infty$.
 Therefore, by making use of Lemma~\ref{Le.3.2}
we get
the limiting relationship (\ref{4.9}) which
implies that $\limsup_\zs{u\to\infty}\,u^\beta\Psi(u)\le C_1$. 
Thus, to finish the proof we need to show 
the inequality $C_1\le C^*(\beta)$.
Indeed, if $0<\beta\le 1$, then \\
$\E((\xi_1+R)^\beta-R ^\beta)\le \E \xi^\beta_1$
 and,
therefore, in this case $C_1\le C^*(\beta)$.
If $\beta>1$, then, taking into account the inequality
$a^\beta-b^\beta\le \beta\,(a-b)\,a^{\beta-1}$ ($0<b<a$),
  we obtain that
$C_1\le 2\alpha\E\xi_1(\xi_1+R)^{\beta-1}/\beta\sigma^2$.
This implies that for $1<\beta\le 2$,
\begin{equation}\label{4.12}
C_1\le 
\frac{2\alpha}{\beta\,\sigma^2}\,
(\E\,\xi^\beta_1+\E\,\xi_1\,\E\,R^{\beta-1})\le 
\frac{2\alpha}{\beta\,\sigma^2}\,
(\E\,\xi^\beta_1+(\E\,\xi^\beta_1)^{\frac{1}{\beta}}\,
\E\,R^{\beta-1})\,.
\end{equation}
Since by  (\ref{4.5}) we have   $\E\,M^{\beta-1}_1\,<\,1$, therefore
by making use of (\ref{4.8}) and taking into account that 
$(\E\,M^{\beta-1}_1)^{-1}-1=\varrho$ ($\varrho$ is defined in (\ref{2.2}))
we can estimate $\E\,R^{\beta-1}$ as
$$
\E\,R^{\beta-1}\le \frac{\E\,Q^{\beta-1}_1}{1-\E\,M^{\beta-1}_1}=
\frac{\E\,\xi^{\beta-1}_1\,\E\,M^{\beta-1}_1}{1-\E\,M^{\beta-1}_1}
\le 
\frac{1}{\varrho}\,(\E\,\xi^\beta)^{\frac{\beta-1}{\beta}}\,.
$$
Thus, from this and (\ref{4.12}), we obtain that 
$C_1\le C^*(\beta)$ for $1< \beta \le 2$.
Let us consider now the case $\beta>2$. In this case 
we estimate $C_1$ as 
\begin{equation}\label{4.13}
C_1\le \frac{2^{\beta-1}\alpha}{\beta\sigma^2}
(\E\xi^\beta_1+\E\xi_1R^{\beta-1})
\le
\frac{2^{\beta-1}\alpha}{\beta\,\sigma^2}
(\E\xi^\beta_1+(\E\xi^\beta_1)^{\frac{1}{\beta}}
\E R^{\beta-1})\,.
\end{equation}
We set
$\|R\|_q=(\E\,R^q)^{\frac{1}{q}}$ with $q=\beta-1$.
Taking into account  that
the random variables $R_1$ and $M_1$ are independent in (\ref{4.8}),
we obtain that 
$$
\|R\|_q\,=\,\|M_1\,R_1\,+\,Q_1\|_q\le
\|M_1\|_q\,\|R_1\|_q\,+\,\|Q_1\,\|_q\,,
$$
i.e. $\|R\|_q\le \|Q_1\|_q(1-\|M_1\|_q)^{-1}
=\|\xi_1\|_q((\|M_1\|_q)^{-1}-1)^{-1}$.
From this, we find 
\begin{align*}
\E\,R^{\beta-1}\le 
\,\left((1+\varrho)^{\frac{1}{\beta-1}}-1\right)^{1-\beta}\,
(\E\,\xi^\beta_1)^{\frac{\beta-1}{\beta}}\,.
\end{align*}
Applying this inequality to (\ref{4.13}), one obtains  
 $C_1\le C^*(\beta)$ for $\beta > 2$. 
This implies Theorem~\ref{Th.2.1}. \endproof

\renewcommand{\theequation}{5.\arabic{equation}} 
\setcounter{equation}{0}
\section{Lower bound for the ruin probability}

In this section we prove Theorem~\ref{Th.2.2}.
First, notice that the identity (\ref{4.3}) implies 
\begin{equation}\label{5.1}
S_n\le S^*_n:=
 \cE_n\,u+\cE_n\sum_{k=1}^n\cE_k^{-1}\,\zeta^*_k\,,
\end{equation}
where $\zeta^*_k=\eta^*_k-\xi_k$ with
$\eta^*_k:=c^*\int_0^{\theta_k}e^{h_\zs{\tau_k}-h_\zs{u+\tau_{k-1}}}
\d u$.
\noindent Therefore, denoting $T^*_u=\inf\{n\ge 1\,:\,S^*_n<0\}$
we obtain 
\begin{equation}\label{5.2}
\Psi(u)=\P(T_u<\infty)\ge \P(T^*_u<\infty)\,,
\end{equation}
for any $u>0$. Setting 
$Q^*_k=(\xi_k-\eta^*_k)/\lambda_k$ 
in (\ref{5.1}), 
we represent $S^*_n$ in the
following form $S^*_n=  \cE_n\,(u-Y^*_n)$,
where $Y^*_1=Q^*_1$ and for $n\ge 2$,
\begin{equation}\label{5.3}
Y^*_n=Q^*_1+M_1Q^*_2+\cdots+
\prod^{n-1}_{j=1}\,M_j\,Q^*_n\,.
\end{equation}
Therefore, for any $u>0$,
\begin{equation}\label{5.4}
 \P(T^*_u<\infty)=\P(R^*>u)\,,
\end{equation}
where $R^*=\sup_{n\ge 1}\,Y^*_n$.
\noindent To study the tail behaviour of $R^*$
we need to obtain the renewal equation for $R^*$.
To this end  
we rewrite $Y^*_n$ as
$Y^*_n=Q^*_1+M_1\,Z^*_n$
with $Z^*_2=Q^*_2$ and 
$Z^*_n=Q^*_2+M_2Q^*_3+\cdots+
\prod^{n-1}_{j=2}\,M_j\,Q^*_n$
for $n\ge 2$. By denoting $R_1:=\sup_{n\ge 2}\,Z^*_n$ 
we get that $R^*=Q^*_1+ M_1\,(R^*_1)_+$. Note that 
the random vector $(Z^*_2,\ldots,Z^*_n)$ has the same
 distribution as $(Y^*_1,\ldots,Y^*_{n-1})$ for any $n\ge 2$, i.e.
 $R^*$ has the same distribution as $R_1$ also.
 Moreover, taking into account that
$R^*_1$ is independent of $(Q^*_1,M_1)$, we deduce 
that $R^*$ satisfies the random Eq. \eqref{3.2}.
We  show now that
\begin{equation}\label{5.6}
\lim_{u \rightarrow\infty}\, u^{\beta}\,\P(R^*>u)= C_*>0\,.
\end{equation}
To prove this we check the conditions of Lemma~\ref{Le.3.3}.
First, notice that (\ref{4.5}) implies (\ref{3.8})
for any $0<\delta<\sqrt{\alpha_1+\beta^2/4}-\beta/2$
with $\alpha_1=2\alpha/\sigma^2$. It easy to see that for such 
$\delta$ and any $x\in\bbr$ in this case
$\E\,M^{\beta+\delta+ix}_1\ne 1$.
Now we verify (\ref{3.10}).
 Writing $q=\beta+\epsilon$
with  $\epsilon>0$, we obtain 
\begin{align*}
\E\,|Q^*_1|^q\le \mbox{const}\,(\E\,M^q_1\,\E\,\xi^q_1+
\E\,(\eta^*_1\,M_1)^q)\,.
\end{align*}
By the conditions of Theorem~\ref{Th.2.1} and (\ref{4.5})
the first term in this inequality is finite for sufficiently
small $\epsilon$. Moreover, we prove that there exists $\varepsilon>0$ such that 
$\E\,(\eta^*_1\,M_1)^q<\infty$.
Indeed, setting  $w^*_\zs{u}=\sup_\zs{0\le s\le u} (- w_\zs{s}-\frac{\kappa}{\sigma}\,s )$ we get
\begin{align*}
\E\,(\eta^*_1\,M_1)^q&=(c^*)^q\,
\E\,\left(\int^{\theta_1}_0\,e^{-\sigma w_\zs{u}-\kappa\,u  }\,
\d u\right)^q\\ 
&\le\,
(c^*)^q\,\E\,\theta^q_1\,e^{q\sigma w^*_\zs{\theta_\zs{1}}}
=(c^*)^q\,\alpha\,
\int^\infty_\zs{0}\,t^q\,\E\,e^{q\sigma w^*_\zs{t}}\,e^{-\alpha t}\,\d t\,.
\end{align*}
The last intergal we estimate as
\begin{align*}
\int^\infty_\zs{0}\,t^q\,\E\,e^{q\sigma w^*_\zs{t}}\,e^{-\alpha t}\,\d t\,\le\,
\frac{2}{\alpha}
\,K^*_\zs{1}\,
\E\,e^{q\sigma w^*_\zs{\tau}}
\,,
\end{align*}
where
$K^*_\zs{1}=\sup_\zs{t\ge 0}(t^q\,e^{-\frac{\alpha}{2}t}) $
and
 $\tau$ is an exponential random variable with the parameter $\alpha/2$ independing
on $(w_\zs{u})_\zs{u\ge 0}$.  Moreover, taking into account that the random variable 
$w^*_\zs{\tau}$ is exponential 
 (see, for example, \cite{BoSa} p. 197) we find that 
$$
K^*_\zs{2}=\E\,e^{q\sigma w^*_\zs{\tau}}=
\frac{\sqrt{\alpha\sigma^2+\kappa^2}+\kappa}{\sqrt{\alpha\sigma^2+\kappa^2}-\kappa-\varepsilon\sigma^2}<\infty
$$
for $0<\varepsilon\sigma^2<\sqrt{\alpha\sigma^2+\kappa^2}-\kappa$.
Therefore we get
\begin{equation}\label{5.8}
\E\,(\eta^*_1\,M_1)^q\le 2(c^*)^q\,K^*_\zs{1}\,K^*_\zs{2}\,.
\end{equation}
Now (\ref{5.6}) follows
from Lemma~\ref{Le.3.3}. Hence Theorem~\ref{Th.2.2}.
\endproof

\renewcommand{\theequation}{6.\arabic{equation}} 
\setcounter{equation}{0}
\section{Exact asymptotics for the ruin probability}

In this subsection we prove Theorem~\ref{Th.2.3}. 
For $\gamma=0$,  the theorem  follows
from (\ref{5.6}).
Therefore we assume $-\infty<\gamma<0$.
 In this case Eq. \eqref{4.2} has the following
form
\begin{equation}\label{6.1}
S_n= \cE_n u+\cE_n\sum_{k=1}^n\cE_k^{-1}(c_{k-1}
\tilde{\eta}_k-\xi_k)\,,
\end{equation}
where
$c_n=c_{\tau_n} = c^* \exp\{\gamma\tau_n\}$
 and
$\tilde{\eta}_n= \int_0^{\theta_n}
e^{h_\zs{\tau_n}-h_\zs{u+\tau_{n-1}}+\gamma u} \d u$.
We set
$\tilde{Y}_n:=
\sum_{k=1}^n\cE_k^{-1}c_{k-1}
\tilde{\eta}_k=
\sum_{k=1}^n\,\prod^{k-1}_{j=1}\,\tilde{M}_j\,
\tilde{Q}_k $
with $\tilde{Q}_k=c^*\,M_k\,\tilde{\eta}_k$
and\\ $\tilde{M}_k=e^{\gamma\theta_k}M_k$. 
Taking into account 
that $(\tilde{Y}_n)$ is an increasing sequence, we put
$$
\tilde{R}=\tilde{Y}_\infty=\lim_{n\to\infty}\,\tilde{Y}_n=
\sum_{k=1}^\infty\,\prod^{k-1}_{j=1}\,\tilde{M}_j\,
\tilde{Q}_k
\quad\mbox{a.s.}
$$
Notice now that this random variable 
satisfies the following identity in law
$$
\tilde{R}\stackrel{(d)}{=} \tilde{Q}+\tilde{M}\tilde{R}\,,
$$
where $\tilde{Q}\stackrel{(d)}{=}\tilde{Q}_1$,
 $\tilde{M}\stackrel{(d)}{=}\tilde{M}_1$ and 
$\tilde{R}$ is independent of $(\tilde{Q},\tilde{M})$.
Moreover, for $q=\tilde{\beta}=\beta-2\gamma/\sigma^2$
we get
$\E\,\tilde{M}^q=
\alpha (\alpha+(\tilde{\beta}-q)\,q\,\sigma^2/2)^{-1}= 1$ and
similarly to (\ref{5.8}) we can show that 
$\E\,\tilde{Q}^{\tilde{\beta}}\,<\,\infty$.
Therefore, Lemma~\ref{Le.3.2} implies that
$\lim_{u \rightarrow\infty}\,
u^{\tilde{\beta}}\,\P(\tilde{R}>u)<\infty$.
Thus, by (\ref{4.9}) 
\begin{equation}\label{6.3}
\lim_{u \rightarrow\infty}\,
\frac{\P(\tilde{R}>u)}{\P(R>u)}=0\,.
\end{equation}
\noindent Now we study the stopping time (\ref{4.1}) in our case.
First,  by (\ref{6.1}) we may write $T_u$ as
\begin{equation}\label{6.4}
T_u:=\inf\{n\ge 1:\ S_n< 0\}=\inf\{n\ge 1:\ Y_n> u+\tilde{Y}_n\}\,,
\end{equation}
where $Y_n$ is defined in (\ref{4.4}).

Recall that, $R=Y_\infty=\lim_{n\to\infty}\,Y_n$ a.s. and
 $\tilde{R}=\tilde{Y}_\infty=
\lim_{n\to\infty}\,\tilde{Y}_n$ a.s..
Therefore from (\ref{6.4}) it follows that 
$\P(R> u+ \tilde{R}\,,\,T_u=\infty)=0$. 
Taking this into account, it easy to deduce the following
equality
\begin{equation}\label{6.5}
\P(T_u<\infty) =  \P(Y_\zs{T_u}>u+\tilde{Y}_\zs{T_u})
 = \P(R>u+\tilde{Y}_{T_u})\,.
\end{equation}
\noindent
 From here we obtain for any $\delta>0$, 
\begin{align*}
\P(T_u<\infty) 
& \ge \P(R>u+\tilde{Y}_\zs{T_u},\,\tilde{Y}_\zs{T_u}\le \delta u)
 \ge \P(R>(1+\delta)u, \tilde{Y}_{T_u}\le
\delta u)\\
& =
\P(R>(1+\delta)u)-\P(R>(1+\delta)u\,,
\tilde{Y}_{T_u}> \delta\,u)\\
& \ge  \P(R>(1+\delta)u)-\P(\tilde{R}>\delta\, u)\,.
\end{align*}
\noindent The limiting
relationships (\ref{4.9}) and (\ref{6.3}) 
imply that
$$
\liminf_{u\rightarrow +\infty}\P(T_u<\infty)/\P(R>u)\ge\,1\,.
$$
Moreover, by (\ref{6.5}) we obtain 
$\P(T_u<\infty)\le \P(R>u)$ for any $u>0$.
Thus
$$
\lim_{u\to\infty}\,\P(T_u<\infty)/\P(R>u)=1\,.
$$
If $\gamma=-\infty$, i.e. $c_\zs{t}=0$,  then
$\tilde{Y}_n=0$ for all $n\in\bbn$ and , hence,
$\P(T_u<\infty)=\P(R>u)$. 
 Therefore (\ref{4.9}) implies this theorem in this case.
\endproof

\renewcommand{\theequation}{7.\arabic{equation}} 
\setcounter{equation}{0}
\section{Erdodic properties for the random coefficient autoregressive
process}

To show Theorem~\ref{Th.2.4} we need to use some ergodic properties of the
special autoregressive process 
with random coefficients (\ref{5.1}).
In this section we study the ergodic properties for a
 general scalar autoregressive process 
with random coefficient
\begin{equation}\label{7.1}
x_n\,=\,a_n\,x_{n-1}\,+\,b_n\,, \ n\ge 1\,,
\end{equation}
where $x_0$ is some fixed constant and $(a_n\,,b_n)$ is i.i.d.
sequence of random variables in $\bbr^2$.

\begin{proposition}\label{Pr.7.1}
Assume that there exists $0<\delta\le 1$ such that $\rho=\E\,|a_1|^\delta<1$
and $\E\,|b_1|^\delta\,<\,\infty$. Then
 for any bounded uniformly continuous function $f$
\begin{equation}\label{7.2}
\P\,-\,\lim_{N\to\infty}\,N^{-1}\,\sum^N_{n=1}\,f(x_n)\,=\,
\E\,f(x_\infty)\,,
\end{equation}
where $x_\infty=\sum^\infty_{k=1}\pi_{k-1}b_k$ with $\pi_0=1$ and
$\pi_k=\prod^k_{j=1}\,a_j$ for $k\ge 1$.
\end{proposition}
\noindent {\bf Proof.}
First we show that the series in the definition of $x_\infty$
 converges in probability.
 Indeed,
$\E\,|\sum^{n+m}_{k=n}\,\pi_{k-1}\,b_k|^\delta\,\le\,\E\,|b_1|^\delta\,
\sum^{n+m}_{k=n}\,\rho^k$.
It means that the series $\sum_{k\ge 1}\pi_{k-1}\,b_k$ converges
in  $\L_\delta$ and hence in probability. 
Now we fixe some $m\ge 1$ and, for $n\ge m$, we set 
$x_n(m)\,=\,\sum^n_{k=n-m+1}\,b_k\,\prod_{j=k+1}^n\,a_j$.
Notice that $x_n(m)$ is mesurable with respect to 
$\sigma\{a_{n-m+1},\ldots,a_n,b_{n-m+1},\ldots,b_n\}$. Therefore
for any $0\le d<m$ the 
the sequence $(x_\zs{km+d})_\zs{k\ge 1}$ is i.i.d. and by 
the  law of large numbers for any fixed $m\ge 1$ and $0\le d<m$
\begin{equation}\label{7.5}
\lim_{p\to\infty}\,p^{-1}\,\sum^p_{k=1}\,f(x_\zs{km+d}(m))\,=\,
\E\,f(x_m(m))\ \ \ \ \ \mbox{a.s.}\,,
\end{equation}
where 
$x_m(m)=\sum^m_{k=1}\,b_k\,\prod_{j=k+1}^m\,a_j\,
 \stackrel{(d)}{=}\,\sum^m_{k=1}\,b_k\,\pi_{k-1}$.
Therefore
\begin{equation}\label{7.6}
\lim_\zs{m\to\infty}\,\E\,f(x_m(m))\,=\,\E\,f(x_\infty)\,.
\end{equation}
We show now that for any $\epsilon>0$
\begin{equation}\label{7.7}
\lim_\zs{m\to\infty}\,\sup_\zs{N\ge m}\,
\P(\,\Delta(N,m)\,>\,\epsilon\,)\,=\,0\,,
\end{equation}
where $\Delta(N,m)=N^{-1}\sum^N_{n=m}\,
|f(x_\zs{n})-f(x_\zs{n}(m))|$.

We put 
$x^*_n(m)\,=x_n-x_n(m)=
\,x_\zs{n-m}\,
\prod^n_{k=n-m+1}\,a_j$. Taking into account that there exists some $L^*<\infty$
such that for any $n\ge 1$
\begin{align*}
\E\,|x_n|^\delta\,&=
\,\E\,|x_0\,\prod^n_{k=1}\,a_j\,+
\sum^{n}_{k=2}\,b_k\,\prod^n_{j=k+1}\,a_j\,|^\delta\\
&\le\,|x_0|^\delta\,\rho^n\,+\,\E\,|b_1|^\delta\,
\sum^{n}_{k=2}\,\rho^{n-k}\le L^*\,,
\end{align*}
we get 
$\sup_\zs{n\ge m}\,
\E\,|x^*_n(m)|^\delta\,\le \,L^*\,\rho^m$.

Let us choose  $\epsilon_1>0$ for which 
$\sup_\zs{|x-y|\le \epsilon_1}\,|f(x)\,-\,f(y)|\,\le \,\epsilon/2$.
For such $\epsilon_1$ we obtain that
$\Delta(N,m)\le
\epsilon/2\,+\,2f^*\,N^{-1}\,\sum^N_{n=m}\,
\Chi_\zs{\{|x^*_n(m)|\ge \epsilon_1\}}$,
where $f^*=\sup_\zs{x\in\bbr}\,|f(x)|$. Therefore by denoting
 $\epsilon^*=\epsilon/4f^*$ we get that
$$
\P(\,\Delta(N,m)\,>\,\epsilon\,)\, \le\,
\P\left(\sum^N_{n=m}\,
\Chi_\zs{\{|x^*_n(m)|\ge \epsilon_1\}}\,>\,\epsilon^*N\right)\,.
$$
 Applying here the Chebyshev inequality we find that
\begin{align*}
\P(\,\Delta(N,m)\,>\,\epsilon\,)\le
\frac{1}{\epsilon^*\,N}\sum^N_{n=m}\,\P(|x^*_n(m)|\ge \epsilon_1)
\le L^*\frac{1}{\epsilon^\delta_1\,\epsilon^*} \rho^m\,.
\end{align*}
This implies (\ref{7.7}).
We put $p=[N/m]$ ($[a]$ is the whole part of $a$), i.e.
 $N=pm+r$ with $0\le r<m$). For such $p$ and $r$, we can write that
\begin{align*}
\Omega_\zs{N}\,:=\,\left|\frac{1}{N}\,\sum^N_{n=1}\,f(x_n)\,-\,
\E\,f(x_\infty)\right|&\,\le\,\left|\frac{1}{N}\,\sum^{pm-1}_{n=m}\,f(x_n(m))\,-\,
\E\,f(x_\infty)\right|\\
&+\,f^*\frac{m+r}{N}\,+\,\Delta(N,m)\,.
\end{align*}
Moreover, we can represent the last sum in this inequality as
$$
\sum^{pm-1}_{n=m}\,f(x_n(m))=\sum^{m-1}_{d=0}\,
\sum^{p-1}_{k=1}\,f(x_{km+d}(m))\,.
$$
Therefore, from (\ref{7.5}), we get that
\begin{align*}
\lim_\zs{N\to\infty}\,\frac{1}{N}\,\sum^{pm-1}_{n=m}\,f(x_n(m))\,&
=\,\lim_\zs{p\to\infty}\,\frac{1}{m}\,\sum^{m-1}_{d=0}\,\frac{1}{p}
\sum^{p-1}_{k=1}\,f(x_{km+d}(m))=\E\,f(x_m(m)) \,.
\end{align*}
Finally, for $\epsilon>0$, we obtain that for any $m\ge 1$
\begin{align*}
\limsup_\zs{N\to\infty}\,\P\left(
\,\Omega_\zs{N}\,>\,\epsilon\,\right)\,\le\,
\sup_\zs{N\ge 1}\,\P\left(
\,|\E\,f(x_\infty)-\E\,f(x_m(m))|\,
+\,\Delta(N,m)>\epsilon\,\right)\,.
\end{align*}
The limiting relationships
(\ref{7.6})--(\ref{7.7}) imply (\ref{7.2}).
\endproof

\renewcommand{\theequation}{8.\arabic{equation}} 
\setcounter{equation}{0}
\section{Large volatility}

In this section we prove Theorem~\ref{Th.2.4}.
 First, notice that if $\beta<0$ then
 Proposition 4 in \cite{FrKaPe} implies that 
$\P(T^*_u<\infty)=1$ for any $u\ge 0$. Thus Theorem~\ref{Th.2.4}
for $\beta<0$ directly follows from Inequality (\ref{5.2}).
  We consider the critical case $\beta=0$, i.e.
$\kappa=0$ and
$\lambda_k=e^{\sigma\nu_k}$
with $\nu_k=w^k_\zs{\theta_k}=w_\zs{\tau_k}-w_\zs{\tau_{k-1}}$.

For this,  we study the ergodic properties
of the process $(S^*_n)$ defined in (\ref{5.1}).
 Notice that
(\ref{5.1}) implies that this process satifies
the following random reccurence equation
\begin{equation}\label{8.1}
S^*_n=\lambda_n\,S^*_{n-1}\,+\,\zeta^*_n\,,
\end{equation}
where $S^*_0=u$ and $\zeta^*_n$ is defined in (\ref{5.1}).

Set $t_0=0$ and 
$t_n=\inf\{k>t_{n-1}\,:\,\sum^k_{j=t_{n-1}+1}\,\nu_j<0\}$ for $n\ge 1$.
It is easy to see that $t_n=\sum^n_{j=1}\,\rho_j$, where
 $(\rho_j)$ is an i.i.d. sequence which has the same 
distribution as $t_1$ 
whose properties are well known,  
see  XII. 7 theorem 1a in \cite{Fe}. One can show,
that for some constant $0<c<\infty$,  
\begin{equation}\label{8.3}
\sup_{n\ge 1}\,n^{1/2}\P(t_1>n)\le c\,.
\end{equation}
\noindent Set $x^*_n=S^*_\zs{t_n}$. By (\ref{8.1}) 
we obtain that for any $n\ge 1$, 
\begin{equation}\label{8.4}
x_n=a_n\,x_{n-1}+b_n\,, \ \ \ x_0=u\,, 
\end{equation}
where $a_n=\prod^{\rho_n}_{j=1}\, \lambda_\zs{t_{n-1}+j} 
=\exp\{\sigma\sum^{\rho_n}_{j=1}\,
\nu_\zs{t_{n-1}+j}\}$
and
$$
b_n=\sum^{\rho_n}_{k=1}\,
(\prod^{\rho_n}_{j=k+1}\, \lambda_\zs{t_{n-1}+j})\,
\zeta^*_\zs{t_{n-1}+k}\,.
$$
The sequence 
$(a_n,b_n)$ is an i.i.d.
sequence of random variables in $\bbr^2$. Moreover, 
$\E\,a_n=\E\,a_1<1$. We will show that there exists  
$r>0$ such that 
\begin{equation}\label{8.5}
\E\,|b_1|^r<\infty\,.
\end{equation}
First, notice that the definition of $b_1$ implies that
$|b_1|\le \sum^{t_1}_{k=1}\,|\zeta^*_k|$. Moreover, similarly
to (\ref{5.8}) we can show that there exists $0<\epsilon<1$
for which $\E\,|\eta^*_1|^\epsilon<\infty$. Therefore, taking into account
the condition of Theorem~\ref{Th.2.4} 
($\E\,\xi^\delta_1<\infty$ for some $\delta>0$) we get that
there exists $0<\epsilon<1$ such that
$m_\epsilon=\E\,|\zeta^*_1|^\epsilon<\infty$.
To finish the proof of inequality (\ref{8.5}),  note that,
 for such $\epsilon$ and for some fixed $0<r<1$,
by making use of inequality (\ref{8.3})  we obtain that
\begin{align*}
\E\,|b_1|^r&\le 1+\,r\,
\sum^\infty_{n=1}\,\frac{1}{n^{1-r}}\,
\P(\sum^{t_1}_{k=1}\,|\zeta^*_k|>n)\\
&\le 1+\,r\,
\sum^\infty_{n=1}\,\frac{1}{n^{1-r}}\,
\P(\sum^{l_n}_{k=1}\,|\zeta^*_k|>n)+\,r\,
\sum^\infty_{n=1}\,\frac{1}{n^{1-r}}\,\P(t_1>l_n)\\
&\le 1+\,r\,m_\epsilon\,
\sum^\infty_{n=1}\,\frac{l_n}{n^{1-r+\epsilon}}\,
+\,r\,c\,
\sum^\infty_{n=1}\,\frac{1}{n^{1-r}l_n^{1/2}}\,.
\end{align*}
Therefore, by putting $l_n=[n^{4r}]$,
 we obtain (\ref{8.5}) for 
$0<r<\epsilon/5$.
Hence, by Proposition~\ref{Pr.7.1}, the process 
(\ref{8.4}) has the property (\ref{7.2}) for
some bounded uniform continuous function $f$.

For Eq. (\ref{8.4}) we reprsent   
the random variable $x_\infty=\sum\zs{k\ge 1}\pi_{k-1}b_k$ as
$x_\infty:=\prod^{t_1}_{j=2}\lambda_j(\varsigma-\xi_\zs{1})$,
where $\varsigma$ is independent of $\xi_1$. 
This implies that $\P(x^*_\infty<0)=\P(\xi_1>\varsigma)$.
Thus, by the condition on the distribution of
$\xi_1$ we obtain that $\P(x^*_\infty<0)>0$. 
It means that for the function
$f_1(x)=\min(x^2,1)\Chi_\zs{\{x\le 0\}}$ we have $\E\,f_1(x_\infty)>0$
and  by (\ref{7.2})  there exists a sequence $(n_k)$ such that\\
$\lim_{k\to\infty}n_k^{-1}\sum^{n_k}_{j=1}\,f_1(x_j)\,=\,
\E\,f_1(x_\infty)>0$ a.s. Therefore $\P(T^*_u<\infty)=1$
and Theorem~\ref{Th.2.3}  follows directly from (\ref{5.2}).
\endproof\\[1mm]

\noindent{\bf Acknowledgments}

The authors thank 
Thomas Mikosch and Claudia Kl\"uppelberg 
 for helpful remarks and comments.


\begin{thebibliography}{100}

\bibitem{As}
S. Asmussen, Ruin Probabilities, World Scientific,
Singapore, 2000.

 \bibitem{BoSa}
A. Borodin, P. Salminen,
Handbook of Brownian Motion and Formulae, 
Birkhauser, 1996.

\bibitem{Dj}
B. Djehiche, A large deviation estimate for ruin
probability. Scand. Actuar. J. 1 (1993) 42--59.

\bibitem{Fe}
W. Feller, An Introduction to Probability Theory and Its Applications,
vol. 2, Wiley, New York, 1966.

\bibitem{FrKaPe} 
A.G. Frolova, Yu.M. Kabanov, S.M. Pergamenshchikov, 
In the insurance business risky investments are dangerous, Finance and Stochastics
 6 (2002) 227--235.

\bibitem{GaGrSc}
J. Gaier, P. Grandits, W. Schachermayer, 
Asymptotic ruin probability and optimal investment,
 Ann. Appl. Probab. 13 (2003) 1054--1076.

\bibitem{Go} 
 C.M. Goldie,
Implicit renewal theory and tails of 
solutions of random equations, Ann. Appl. Probab. 1 (1991) 126--166.


\bibitem{Ja}
J. Jacod, Calcul stochastique et probl\`emes,
Lecture Notes in Mathematics, vol. 714. Springer, 
Berlin, Heidelberg, New York, 1979.

\bibitem{KaNo} 
V. Kalashnikov, R. Norberg, Power tailed ruin
probabilities in the presence of risky investments,
Stochastic Process. Appl. 98 (2002) 211--228. 

\bibitem{KlKyMa}
C. Kl\"uppelberg, A.E. Kyprianou, R.A. Maller,
Ruin probability and overshoods for general L\'evy processes,
Ann. Appl. Probab. 14 (2004) 1766--1801. 


\bibitem{MaSu}
J. Ma, X. Sun,  Ruin probability for insurance models involving investments,
Scand. Actuar. J. 3 (2003) 217--237

\bibitem{Mi}
T. Mikosch,
 Non-Life Insurance Mathematics,
An Introduction with Stochastic Processes,
Springer, Berlin, 2004.

\bibitem{Ny} 
H. Nyrhinen, Finite and infinite time
 ruin probabilities in a
stochastic economic environment, 
Stochastic. Process. Appl.
 92 (2001) 265--285.


\bibitem{Pa} 
J. Paulsen, Sharp conditions for certain ruin in
a risk process with stochastic return on investments,
Stochastic. Process. Appl. 75 (1998) 135--148.


\end{thebibliography}
\end{document}